\documentclass[letterpaper,english,11pt]{article}

\usepackage{verbatim}
\usepackage{url}
\usepackage{graphicx}
\usepackage{microtype}
\usepackage{geometry} % remove showframe
\usepackage{tabularx}
\usepackage[table]{xcolor}
\usepackage{hyperref}
\usepackage{natbib}

\usepackage{setspace}

\usepackage{subcaption}

\usepackage{soul} % http://ctan.org/pkg/soul

\usepackage[export]{adjustbox}

\usepackage{titlesec}

\titleformat{\section}
  {\centering\sffamily\bfseries\fontsize{14pt}{\baselineskip}\selectfont}{}{0pt}{}
\titlespacing*{\section}{0pt}{13pt}{3pt}
\titleformat{\subsection}% \large
  {\centering\sffamily\bfseries\itshape\fontsize{14pt}{18pt}\selectfont}{}{0pt}{}
\titlespacing*{\subsection}{0pt}{10pt}{3pt}
\titleformat{\subsubsection}
  {\centering\normalfont\scriptsize}{}{0pt}{}
\titlespacing*{\subsubsection}{0pt}{12pt}{12pt}
\DeclareCaptionFormat{custom}
{%
    {\normalfont\normalsize#1#2}{\normalfont\normalsize#3}
}
\DeclareCaptionLabelSeparator{custom}{. }

\makeatletter

\@ifundefined{pageheight}{\let\pageheight\pdfpageheight}{}
\@ifundefined{pagewidth}{\let\pagewidth\pdfpagewidth}{}
\pageheight\paperheight
\pagewidth\paperwidth

\let\SF@@footnote\footnote
\def\footnote{\ifx\protect\@typeset@protect
    \expandafter\SF@@footnote
  \else
    \expandafter\SF@gobble@opt
  \fi
}
\expandafter\def\csname SF@gobble@opt \endcsname{\@ifnextchar[%]
  \SF@gobble@twobracket
  \@gobble
}
\edef\SF@gobble@opt{\noexpand\protect
  \expandafter\noexpand\csname SF@gobble@opt \endcsname}
\def\SF@gobble@twobracket[#1]#2{}

\usepackage{hyperref}

\newcommand{\doi}[1]{{doi:~\href{https://doi.org/#1}{\nolinkurl{#1}}}\rmFullStop}

\newcommand*{\rmFullStop}{\rmifnextchar{.}{}{}}

\makeatletter
\newcommand{\rmifnextchar}[3]{%
  \begingroup
  \ltx@LocToksA{\endgroup#2}%
  \ltx@LocToksB{\endgroup#3}%
  \ltx@ifnextchar{#1}{%
    \def\next{\the\ltx@LocToksA}%
    \afterassignment\next
    \let\scratch= %
  }{%
    \the\ltx@LocToksB
  }%
}
\makeatother

\makeatother

\usepackage{enumitem}
\setlist[itemize, 1]{topsep=0pt}
\setlist[enumerate, 1]{topsep=0pt}

\hypersetup{
	colorlinks=true,
    linkcolor=blue,
    filecolor=blue,
    urlcolor=blue,
    citecolor=blue,
    pdftitle={Ontological Definition of Seamless Digital Engineering Based on ISO/IEC 25000-Series SQuaRE Product Quality Model},
    pdfauthor={James S. Wheaton and Daniel R. Herber},
}

\usepackage{ifthen}
\usepackage{fancyhdr}
\AtBeginEnvironment{table}{%
  \fontsize{7.5\p@}{10.5\p@}\selectfont%
  \rowcolors*{3}{black!10}{}%
  \arrayrulecolor{black!20}%
  \setlength{\arrayrulewidth}{1\p@}%
}

\usepackage{threeparttable}

\usepackage{siunitx}
\usepackage{multirow}
\usepackage{graphicx}
\usepackage{xcolor}
\usepackage{listings}
\usepackage{listings-owl}

\definecolor{nicegray}{HTML}{818181}

\definecolor{CSUgreen}{HTML}{1E4D2B}
\definecolor{CSUorange}{HTML}{D9782D}
\begin{document}

\pagestyle{fancy}

% \begin{center}\sffamily\bfseries\fontsize{18pt}{21pt}\selectfont
% Ontological definition of seamless digital engineering based on ISO/IEC 25000-series SQuaRE product quality model
% \end{center}

\begin{center}\sffamily\bfseries\fontsize{16pt}{19pt}\selectfont
Ontological Definition of Seamless Digital Engineering Based on ISO/IEC 25000-Series SQuaRE Product Quality Model
\end{center}

\begin{center}%
\scshape A Preprint
\end{center}%

\vspace{-\baselineskip}

\begin{center}
\newcolumntype{C}{>{\centering\arraybackslash}X} % centered version of 'X' columns
\noindent
\begin{tabularx}{0.99\textwidth}{CCC}
James S. Wheaton & Daniel R. Herber \\ 
Colorado State University & Colorado State University \\
\href{mailto:james.wheaton@colostate.edu}{james.wheaton@colostate.edu} & \href{mailto:daniel.herber@colostate.edu}{daniel.herber@colostate.edu} \\
\end{tabularx}
\end{center}

\makeatletter
\AtEndEnvironment{lstlisting}{\xdef\xlang{\lst@language}}
%\AfterEndEnvironment{lstlisting}{\begin{flushright}\vspace{-2ex}\scriptsize\color{nicegray}\textit{\ifboolexpr{0 = \pdfstrcmp{\xlang}{manchester-syntax}}{OWL 2 Manchester Syntax}{\xlang}}\end{flushright}}
\AfterEndEnvironment{lstlisting}{\begin{flushright}\vspace{-4ex}\scriptsize\color{nicegray}\textit{OWL 2 Manchester Syntax}\end{flushright}}
\makeatother

\renewcommand\lstlistingnamestyle{\small}
\lstset{%
    language=Manchester-syntax,
    numbers=none,
    %basicstyle=\normalsize
    }%

% \pagenumbering{gobble} % no page numbers

\newcommand{\myabstract}[1]{\textbf{\large Abstract.}~{\normalsize#1}}

\myabstract{%
Since the introduction of Digital Engineering (DE) as a well-defined concept in 2018, organizations and industry groups have been working to interpret the DE concepts to establish consistent meta-models of those interrelated concepts for integration into their DE processes and tools.
To reach the breadth and depth of DE concept definitions, the interpretation of international standard sources is necessary, including ISO/IEC/IEEE 15288, 24765, 42000-series, 15408, 15206, 27000-series, and 25000-series, to effectively model the knowledge domain where digital engineering applies.
The harmonization of the concepts used in these international standards continues to improve with each revision, but it may be more effectively accomplished by relying on the descriptive logic formalized in the Web Ontology Language (OWL 2 DL).
This paper presents a verified and consistent ontology based on the Basic Formal Ontology (BFO) and Common Core Ontologies (CCO) that defines Seamless Digital Engineering as a digital tooling paradigm that relies on formal verification of digital interfaces to provide a system-level qualification of the assured integrity of a Digital Engineering Environment.
The present work defines classes and equivalence axioms, while using only the BFO- and CCO-defined object properties that relate them, to provide a baseline analysis that may inform future DE-related ontology development, using a case study to formally define the `seamless' quality in relation to the updated ISO 25010 SQuaRE product quality model.
We identified ISO meta-model inconsistencies that are resolvable using the BFO/CCO ontological framework, and define `seamless' as both a system integration quality and a Human-Computer Interface quality-in-use, working to disambiguate this concept in the context of DE.%
}

\newcommand{\keywords}[1]{\textbf{\large Keywords.}~{\normalsize#1}}


% \clearpage
\section{Introduction}
\label{sec:intro}

Since the introduction of Digital Engineering (DE) as a well-defined concept in 2018, enterprises and working groups have been working out how to transform existing Model-Based Engineering (MBE) and Digital Twin / Digital Thread strategies and architectures to encompass the whole system lifecycle as envisioned by the DE strategy \citep{dod2018digital}.
Many challenges exist, including tooling capability, integration capability, process digitalization, data and metadata management, and adaptation of existing Digital Transformation (DT) programs to accommodate the DE vision.
Tool vendors and standards committees are actively working to fill the void of unmet needs when it comes to an organization implementing their own DE strategy.
We are now in an exciting era of development for model-based systems engineering (MBSE) and DE that is supported by professional societies, standards bodies, tool vendors, and engineering organizations, so we may soon be practicing DE and realizing its many promised benefits.

As engineers, we rely on precise, prescriptive language to carry out our work, and DE is no exception. Position papers and reference models on Digital Twin / Digital Thread have been published that further describe these concepts in the context of DE \citep{aiaa2020digital, aiaa2023digitalTwin, aiaa2023digitalThread}, along with explication of the concept of Authoritative Source of Truth (ASoT) \citep{Allison2023}.
While these efforts are important steps in understanding and implementing DE, we now have multiple definitions of DE concepts that may overlap or conflict.
Examples include `Digital Engineering Ecosystem' and `Digital Engineering Environment' which may have been used interchangeably but only DE Ecosystem is defined by the DAU Glossary \citep{DAU-Glossary}, while the INCOSE Systems Engineering Handbook discusses the DE Ecosystem concept as dependent on scope and context, without mentioning DE Environment \citep{INCOSE_SE_HB2023}.
To avoid confusion during DE transformation efforts, and to avoid imprecision of data in production environments --- to realize ASoTs --- ontologies have been identified as a critical component of MBE and DE strategies \citep{Dunbar2023}.

To address the integration and systems security challenges of DE owing to its reliance on insolvent \citep{krasner2022cost,becker2023insolvent} digital infrastructure, Seamless DE was proposed by the authors as a grand challenge in DE research \citep{Wheaton2024a}
Following the clean-slate approach first proposed by DARPA \citep{DARPA_CRASH}, Seamless DE aims to provide an alternative path to the brittle integration of DE Environments, instead focusing on end-to-end formal verification, metamodel and metalanguage coherence, and unified Human-Computer Interface to meet the needs of DE practitioners.
This clean-slate approach to DE has the potential to stimulate DE research based on first principles, while taking advantage of recent advancements in MBSE and computer science to deliver real synergy of capabilities unattainable by traditional digital integration methods. To clarify these ideas in the context of DE, an ontological development approach is taken and the mature ISO/IEC/IEEE 25000-series SQuaRE quality and related standards are used to elucidate the meaning and potential of Seamless DE.

The contributions of this paper include: 1) ontological definition of `seamless' and Seamless Digital Engineering in the context of systems engineering standards and based on established top- and mid-level ontologies and the ongoing work of the INCOSE Digital Engineering Information eXchange (DEIX) Ontology Working Group, 2) assimilation and presentation of ISO/IEC/IEEE standards information including the ISO/IEC 25000-series on quality into an ontology used to define these seamless DE concepts, and 3) an open-source, machine-readable Semantic Web standards-based ontology for DE and MBSE applications.

Section~\ref{sec:background} provides additional background on ontology engineering in the context of DE. In Section~\ref{sec:Methodology}, we outline the methodology used to develop the present ontology. Section~\ref{sec:results} presents the primary results, including ontological definitions for `Seamless Quality Claim', `Seamless Integration', `Seamless Interaction Capability', `Seamless Quality-in-Use', `Seamless Digital Engineering Paradigm', and `Seamless Digital Engineering Environment'. Section~\ref{sec:conclusions} concludes and outlines future work.

% Section 5 discusses the potential applications toward digital quality engineering, and the challenges of developing the ontology based on standards definitions and concepts.

% \clearpage
\section{Background}
\label{sec:background}

\subsection{Ontology Engineering}

Ontologies are formal representations of knowledge domains, wherein concepts are defined through their relations and thereby disambiguated. 
Ontologies have the potential to link together adjacent domains of knowledge to facilitate the interchange of structured information among domain experts.
While many ontology languages exist, the mathematical decidability of an ontology is an important contributor to its usefulness.
Based on the axioms provided in a given ontology, inferences of knowledge may be made with a suitable reasoner, and structured queries may be made that give deterministic results.
Overall, ontologies are suitable for providing the accuracy and reliability that engineers need when navigating complex, emerging knowledge domains \citep{gomez2006ontological}.

The Prot{\'e}g{\'e} open-source software tool developed by Stanford University \citep{Musen2015} has proven itself an indispensable tool in ontology engineering work. Prot{\'e}g{\'e} is based on the Web Ontology Language (OWL) version 2 standardized by the World Wide Web Consortium (W3C) \citep{OWL2Primer} as a Semantic Web technology. By using OWL 2, Prot{\'e}g{\'e} interoperates with the digital ecosystem of Semantic Web technologies, including Resource Description Framework (RDF), Extensible Markup Language (XML), and SPARQL Protocol and RDF Query Language (SPARQL).
These open standards enable organizations to use existing software libraries and to write their own tools to handle Semantic Web data for their DE Environments.
Similar to how tool vendors implement the XML-based standards published by the Object Management Group (OMG), Prot{\'e}g{\'e} provides a familiar Graphical User Interface (GUI) for authoring an ontology and importing other ontologies from the Semantic Web. One need only to open an ontology file and begin navigating its entities and relations to familiarize oneself with the tool's capabilities --- ontology imports are handled automatically by the tool to provide a cohesive view of the `imports closure'.

\subsection{Description Logic}

The mathematical formalism offered by OWL 2 and Prot{\'e}g{\'e} is known as a Description Logic (DL), which is a subset of First-Order Logic using propositional logic that ensures its decidability \citep{Grau2008, DLHandbook2007}. The particular DL used by OWL DL subset is characterized by $\mathcal{SHOIN}(\mathbf{D})$, summarized in Table~\ref{tab:SHOIN}.

\begin{table*}[bt]
\centering
\caption{OWL DL summary of symbols of logic and their descriptions}
\label{tab:SHOIN}

\begin{tabularx}{\columnwidth}{>{\centering}p{0.5in} X}
\hline
\textbf{Symbol} & \textbf{Description} \\
\hline
$\mathcal{AL}$ & (Attributive Language) Inclusion, equivalence, intersection, and complex definition of classes \\
$\mathcal{ALC}$ & (with Complement) Adds to $\mathcal{AL}$ the empty, complement, union classes \citep{baader2007description} \\
$\mathcal{S}$ & Adds the transitivity of relations to $\mathcal{ALC}$ \\
$\mathcal{H}$ & Inclusion and equivalence between relations \\
$\mathcal{O}$ & (One of) Classes created with list of all and only the individuals contained
\\
$\mathcal{I}$ & (Reverse) Inverse property \\
$\mathcal{N}$ & (Number) Cardinality restriction \\
$\mathcal{D}_n$ & (Countable domain) Definition of domains (data types) \\
\hline  % Please only put a hline at the end of the table
\end{tabularx}
% \end{threeparttable}
\vspace{\baselineskip}
\end{table*}

In short, OWL 2 ontologies consist of classes, object properties, annotation properties, individuals, and data properties, and are checked by OWL reasoners which implement OWL DL or other profiles described in \citet{OWL2Primer}. Individuals are instances of classes and inherit their subclass and equivalence axioms, enabling object property assertions with other individuals. Individuals provide the capability to ground the ontology in familiar real-world objects, while additionally checking the consistency of the ontology. One example of an individual in our ontology would be Object Management Group (OMG), which has type class `Standards Organization' and `is carrier of' Systems Modeling Language (SysML), an individual of type class `Architecture Description Language'. OWL DL can be represented in the Manchester syntax \citep{OWL2Manchester}, which provides a human- and machine-readable syntax for authoring axioms.

\subsection{Top-Level and Mid-Level Ontologies}

While the development of an ontology in isolation provides the benefits afforded by the DL and the Semantic Web digital tool ecosystem, those benefits are enhanced by importing existing ontologies that model higher-level concepts and relations. The Basic Formal Ontology (BFO) \citep{BFO2020,ISO21838-2} provides the top-level ontology of our domain ontology, including 36 classes based on the principles of ontological realism, fallibilism, and adequatism. The BFO class hierarchy starts with `entity' and is subdivided into `continuant' and `occurrent', i.e., entities that have a continuous existence and entities which are processes and other temporally-bounded entities. Continuants are divided into `generically dependent continuants', `independent continuants', and `specifically dependent continuants'. These types of continuants and occurrents form the basis of the realistic classes of over 350 ontologies. It is important to note that although BFO is based on the principle of realism (entities exist, have existed, or will exist), it does not preclude certain types of modeling in digital engineering of entities that do not exist, such as in simulations, because the digital objects nevertheless exist with a material basis of the constituent computer components, and the target objects are also supported by generically dependent continuants such as Information Content Entities (ICEs).

The Common Core Ontologies (CCO) \citep{CommonCore} is a mid-level ontology suite based on BFO. It contributes eleven ontologies that extend the usability of BFO while remaining domain-agnostic. These include: Extended Relations, Units of Measure, Time, Quality, Information Entity, Geospatial, Facility, Event, Currency Unit, Artifact, and Agent. Among these, the Extended Relation, Quality, Information Entity, Artifact, and Agent ontologies have been immensely useful in defining the Seamless Digital Engineering domain ontologies.

% \clearpage
\section{Methodology}
\label{sec:Methodology}

The present ontology was developed following the Noy and McGuinness methodology presented in \citep{noy2001ontology}. Starting with our DE knowledge domain, BFO and CCO top- and mid-level ontologies, we begin by identifying concepts and defining classes within the existing class hierarchy. We then move on to defining relations or object properties, and instances of the classes to further validate the ontology. The steps of the \citet{noy2001ontology} methodology, adapted to OWL terminology, are as follows:

\begin{enumerate}[nosep]
    \item Determine the domain and scope of the ontology
    \item Consider reusing existing ontologies
    \item Enumerate important terms in the ontology
    \item Define the classes and the class hierarchy
    \item Define the object properties
    \item Define the domains/ranges of object properties
    \item Create instances of classes
\end{enumerate}

The process becomes iterative as the reasoner validates each step, and missing concepts are identified.
We are using the HermiT 1.4 reasoner \citep{mgs08structured-objects} to check the consistency of the ontology. When a change is made to an `equivalent to' or `subclass of' axiom, the reasoner is synchronized with the ontology and checks its consistency. 
This way, erroneous definitions are caught immediately and flagged as \texttt{owl:Nothing} inferences, which may then be explained using the reasoner's capabilities.

To be a valid axiom accepted by the Prot{\'e}g{\'e} software and its reasoners, every class and object property used in an axiom must already have a valid definition.
Ontology development proceeds iteratively in this way until all the associated concepts reach satisfactory and stable definitions.
Refactoring may then proceed with automated validation, checking each change to maintain the consistency of the ontology. This process of adding new classes in the BFO and CCO class hierarchy requires careful consideration of how a new class follows the conventions and axioms. Without these top- and mid-level hierarchies, the domain ontology can easily lack rigor, which adversely affects its reusability and reproducibility.

Domains and ranges of object properties are especially important in this process, as they restrict the possible classes and provide the consistency checking we need to develop the ontology. At this stage of development, the addition of new object properties (Steps 5-6) was avoided to ensure the BFO and CCO were fully used. After careful consideration of the class hierarchy and use of existing object properties, more precise relations may be identified and developed to clarify the meaning of some concepts. However, complex hierarchies of object properties should be avoided to simplify reasoning. 

\begin{figure}[bt]
    \centering
    \includegraphics[width=\columnwidth]{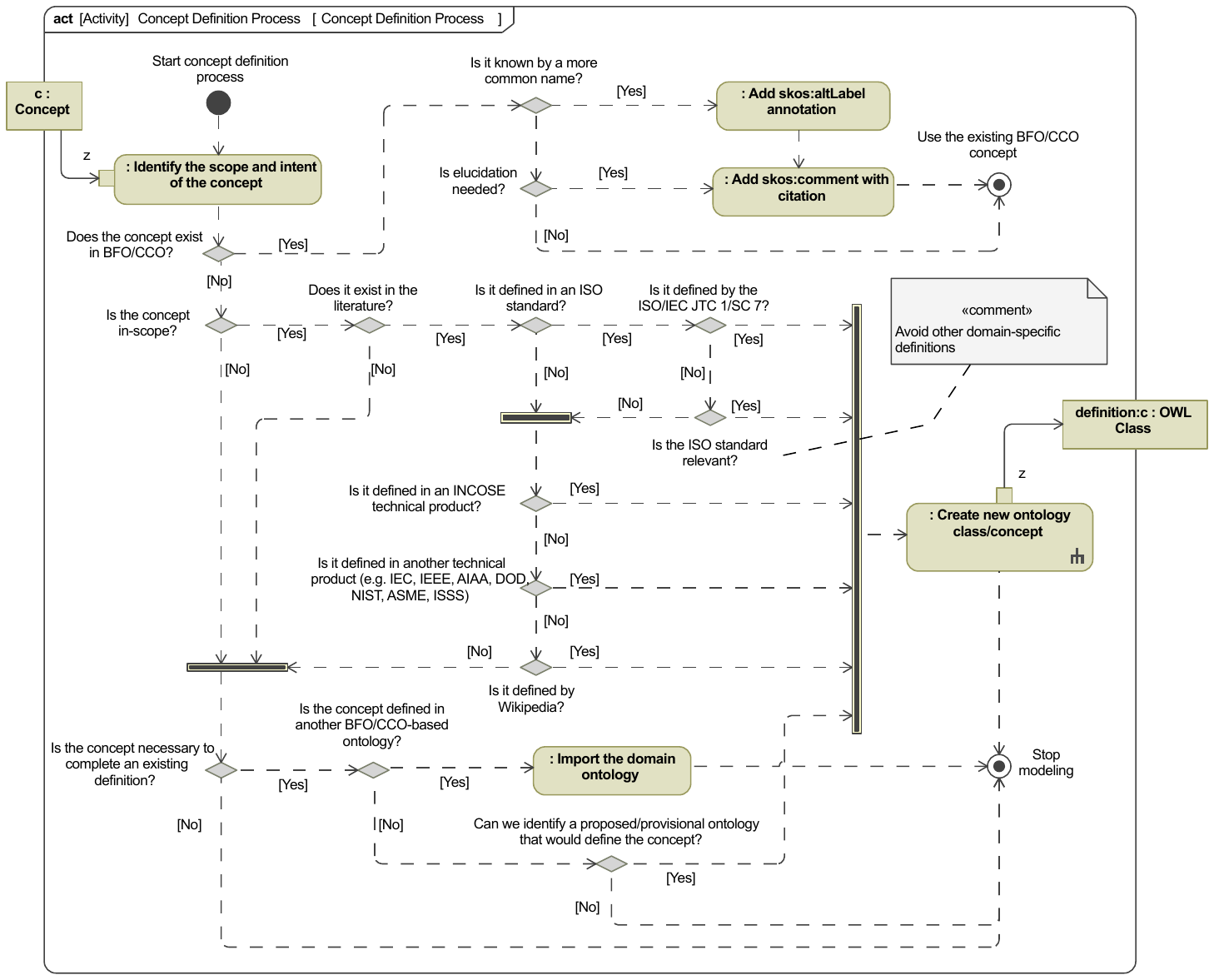}
    \caption{Ontology concept definition process using standard sources}
    \label{fig:ConceptDefinitionProcess}
\end{figure}

To develop the concepts, we referenced international standards, INCOSE technical products, and other canonical sources, but did not reuse any existing DE domain ontology. By referencing international standards in the systems engineering and related domains, we were able to confidently build from recognized meanings of common concepts. However, following the process in Fig.~\ref{fig:ConceptDefinitionProcess} and the practice demonstrated by CCO, when a concept belongs to a more general domain and is not defined by ISO, it often suffices to use Wikipedia or an English dictionary as the source. This concept definition process prioritizes internationally-recognized terminology while leaving room for emerging concepts undergoing consensus definition processes, such as those in DE.

%%%%%
%%%%%

% \clearpage
\section{Results}
\label{sec:results}

The Seamless Digital Engineering Ontology \citep{SDEontologyRepo} consists of over 500 classes and over 150 `equivalent to' axioms, while adhering to a strict zero new object properties defined to ensure logical consistency with BFO and CCO at this stage of development. Definitions related to `seamless' are based on the SQuaRE product quality and quality-in-use models, which are modeled as realizable entities (subclass of specifically dependent continuant) in the ontology. While the BFO provides the `quality' class, it differs from `realizable entity` in that qualities need no further processes to be realized. The quality characteristics according to the standard are defined as capabilities and therefore modeled as capabilities with relations to Information Quality Entities according to CCO. 

Concepts from the following standards and other canonical sources (Table~\ref{tab:Sources}) were used in the development of the Seamless Digital Engineering ontology, reusing BFO and CCO concepts wherever possible to avoid conflicts and duplication.
The ``Count'' column indicates the number of concepts whose definition derives from that source. Notably, the highest count is for concepts defined in \citet{ISO15288}. Although many attempts at developing a systems engineering domain ontology are documented in the literature \citep{gregory2024towards, yang2019ontology}, a recognized standard does not yet exist. Lacking a publicly available systems engineering domain ontology developed in OWL using BFO and CCO, these ontology development efforts were reproduced for this research. The next largest share of definition sources are the ISO/IEC 25000-series SQuaRE product quality standards \citep{ISO25000, ISO25002, ISO25010, ISO25019}, owing to the number of product quality and quality-in-use characteristics defined therein.
\nocite{ISO15026-1,ISO15026-2,ISO15026-4,ISO24765,ISO24748-1,ISO24748-6,ISO21841,ISO24641,ISO26514,ISO42010,ISO42020,ISO15408-1,ISO15408-5,ISO41062,ISO56000,ISO9000,ISO5127,ISO38500,ISO12207,ISO10795,ISO10303-2}

%%%%%
\begin{table*}[tb]
\centering
\caption{Sources used to define ontology classes/concepts (see \protect\hyperlink{sec:References}{References})}
\label{tab:Sources}
% \begin{threeparttable}
% \begin{tabularx}{m{0.18\textwidth}m{0.8\textwidth}}
%\begin{tabularx}{\columnwidth}{>{\small}X >{\small}X >{\small}X >{\small\raggedleft\arraybackslash\hsize=0.2\hsize}X}
\footnotesize
\begin{tabular}	{lllr}
\hline
\textbf{Standard Identifier} & \textbf{Domain} & \textbf{Subject Area} & \textbf{Count} \\ \hline
ISO/IEC/IEEE 15288 & Systems and software engineering & System life cycle processes & 82\\
ISO/IEC 25000 & Systems and software engineering & Guide to SQuaRE & 3\\
ISO/IEC 25002 & Systems and software engineering & SQuaRE --- Quality model overview and usage & 14\\
ISO/IEC 25010 & Systems and software engineering & SQuaRE --- Product quality model & 50\\
ISO/IEC 25019 & Systems and software engineering & SQuaRE --- Quality-in-use model & 13\\
ISO/IEC/IEEE 15026 & Systems and software engineering & Systems and software assurance & 29\\
ISO/IEC/IEEE 24765 & Systems and software engineering & Vocabulary & 18\\
ISO/IEC/IEEE 24748 & Systems and software engineering & Life cycle management & 8\\
ISO/IEC/IEEE 21841 & Systems and software engineering & Taxonomy of systems of systems & 4\\
ISO/IEC/IEEE 24641 & Systems and software engineering & Methods and tools for MBSSE & 1\\
ISO/IEC/IEEE 26514 & Systems and software engineering & Design and development of information for users & 2\\
ISO/IEC/IEEE 42010 & Software, systems and enterprise & Architecture description & 8\\
ISO/IEC/IEEE 42020 & Software, systems and enterprise & Architecture processes & 8\\
ISO/IEC 15408 & Information security and cybersecurity & Evaluation criteria for IT security & 27\\
ISO/IEC/IEEE 41062 & Software engineering & Software acquisition & 3\\
ISO 56000 & Innovation management & Fundamentals and vocabulary & 2\\
ISO 9000 & Quality management systems & Fundamentals and vocabulary & 2\\
ISO 5127 & Information and documentation & Foundation and vocabulary & 6\\
ISO/IEC 38500 & Information technology & Governance of IT for the organization & 2\\
ISO/IEC/IEEE 12207 & Systems and software engineering & Software life cycle processes & 1\\
ISO 10795 & Space systems & Programme management and quality & 3\\
ISO 10303-2 & Industrial automation systems & Product data representation and exchange & 1\\
\hline
\\
\hline
\multicolumn{3}{l}{\textbf{Other Canonical Source}} & \textbf{Count} \\ \hline
\multicolumn{3}{l}{INCOSE Systems Engineering Handbook v5 \citep{INCOSE_SE_HB2023}} & 4\\
\multicolumn{3}{l}{INCOSE Needs and Requirements Manual \citep{NRM2022,GtWR2023}} & 13\\
\multicolumn{3}{l}{Systems Engineering Body of Knowledge \citep{SEBoK}} & 5\\
\multicolumn{3}{l}{NASA Systems Engineering Handbook \citep{NASA_SE_HB2017}} & 14\\
\multicolumn{3}{l}{Defense Acquisition University Glossary \citep{DAU-Glossary}} & 7\\
\multicolumn{3}{l}{Wikipedia} & 11\\
\multicolumn{3}{l}{Academic literature} & 4\\
\multicolumn{3}{l}{Dictionaries \citep{wordnet}} & 6\\
\hline  % Please only put a hline at the end of the table
\end{tabular}
% \end{threeparttable}
\vspace{\baselineskip}
\end{table*}
%%%%%

\begin{figure}[p]
    \centering
    \includegraphics[width=0.87\columnwidth]{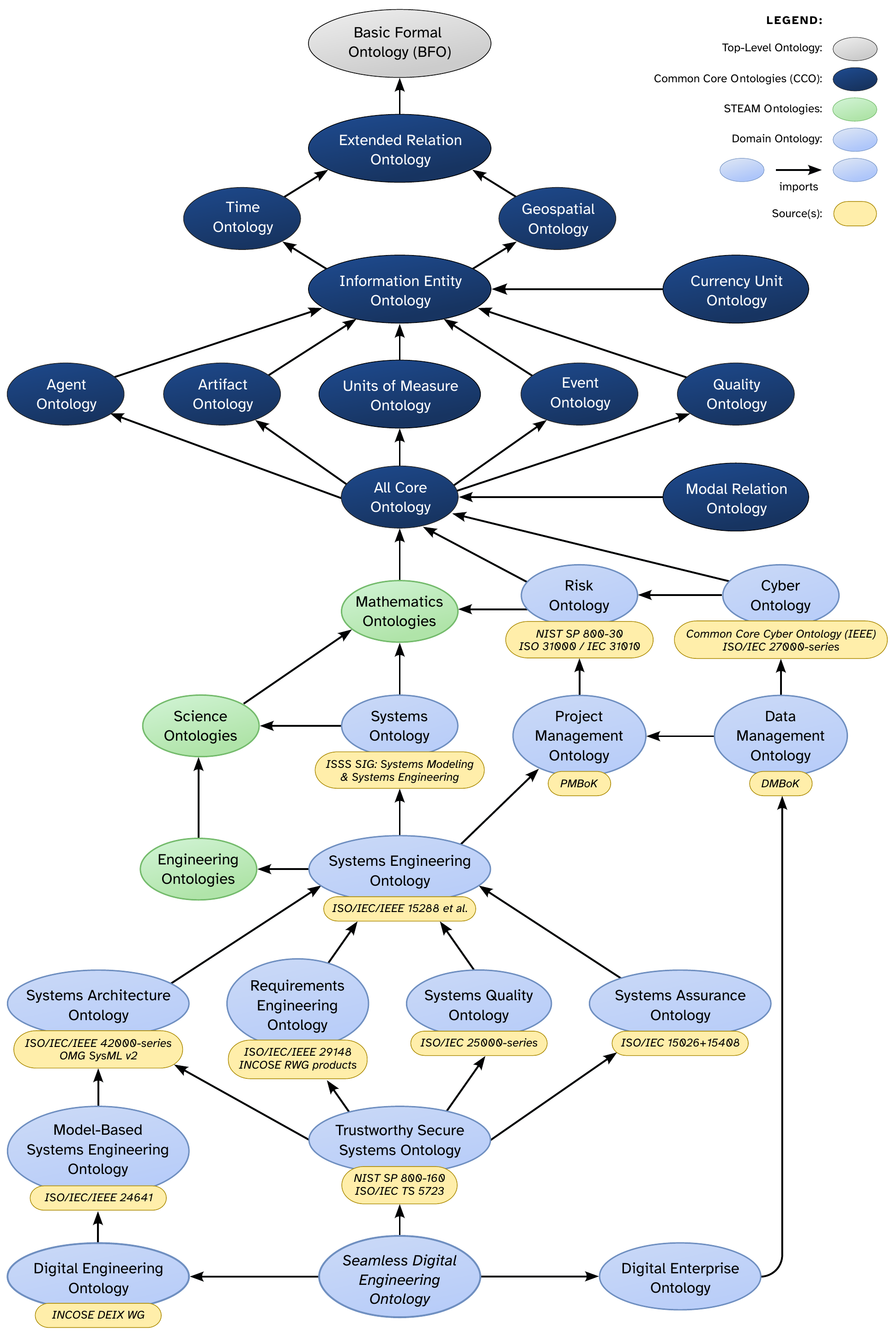}
    \caption{Proposed import hierarchy of DE-related domain ontologies, based on BFO and CCO}
    \label{fig:SDEImportHierarchy}
\end{figure}

The diverse domains and subject areas in Table~\ref{tab:Sources} were drawn from to support the definition of the concepts essential to DE in general, and to Seamless DE in particular. Ideally, these domains would be represented in importable CCO-based ontologies of their own, supporting separation of concerns and reusability. Systems engineering, for example, would be its own domain ontology imported by the DE ontology and would have sub-domain ontologies such as for requirements engineering, systems architecture, and quality engineering.
Figure~\ref{fig:SDEImportHierarchy} depicts a non-exhaustive proposed import hierarchy, and shows that DE relies on concepts in trustworthy secure systems, model-based systems and software engineering (MBBSE), and digital enterprise, including the extensive topics described by the Data Management Body of Knowledge (DMBoK) \citep{DMBoK}.
Such breadth and depth enable precise knowledgebase queries, Authoritative Source of Truth consistency evaluators, and support system architecture modeling activities, including of the Seamless DE Environment system-of-interest.

%%%%%
%%%%%

\begin{figure}[bt]
    \centering
    \includegraphics[width=0.9\columnwidth]{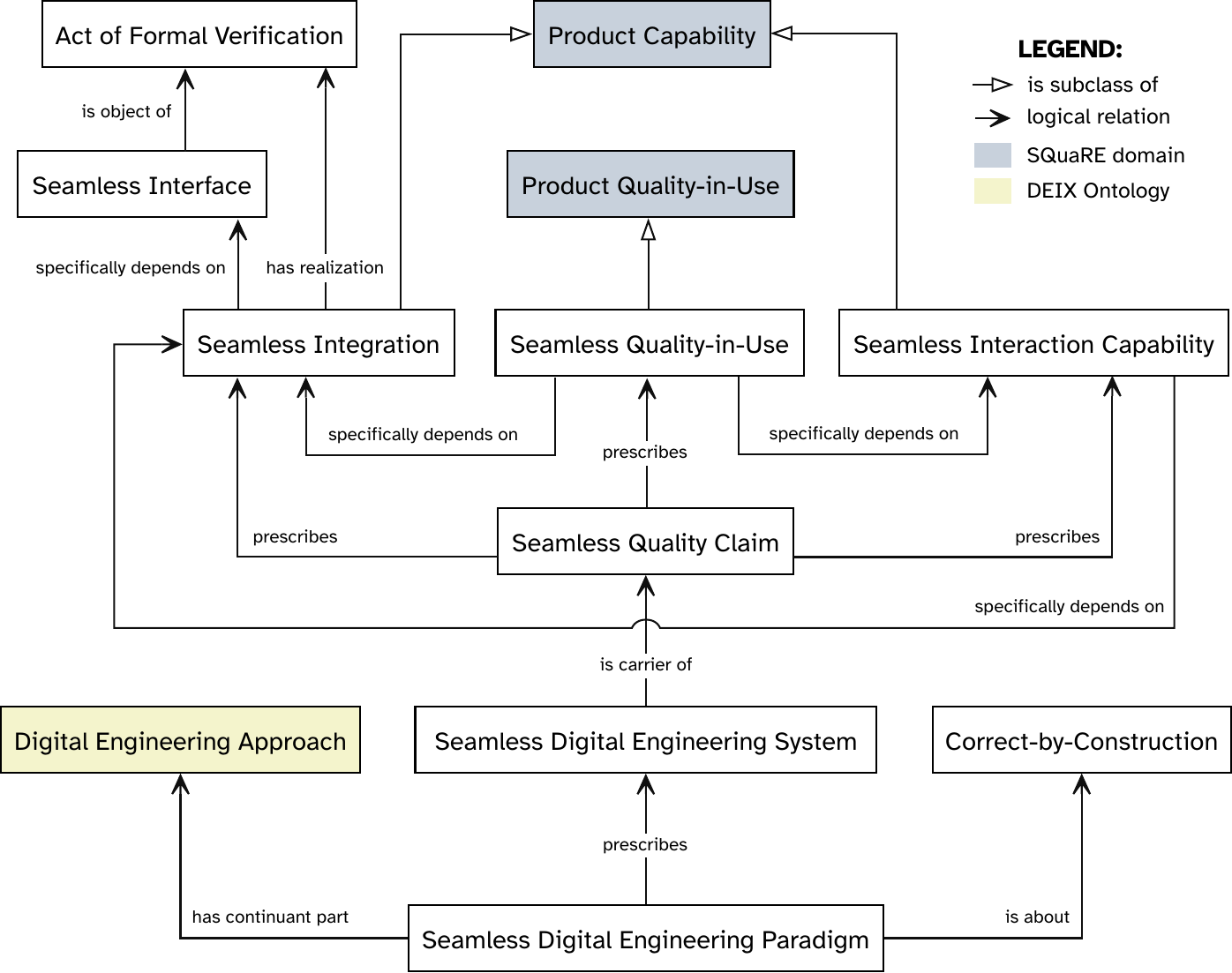}
    \caption{Summary of ontological relations of Seamless Digital Engineering concepts}
    \label{fig:RelationalDiagram}
\end{figure}

The following listings provide the description logic axioms that relate the Seamless DE concepts, including those from the SQuaRE product quality and quality-in-use models \citep{ISO25010,ISO25019}. A summary relational diagram is shown in Fig.~\ref{fig:RelationalDiagram} that highlights how the salient `seamless' concepts relate to one another. The SQuaRE concepts are at the top-center, and not all concepts and relations are shown. Refer to the listings below for more details.

%%%%%
%%%%%

Seamless Digital Engineering was identified as a DE tooling paradigm by \citet{Wheaton2024a}, where the theory developed from related work in `seamless' model-driven systems engineering \citep{Broy2020seamless, broy2010seamless, broy2009seamless}.
Further development of these concepts using an ontological framework and reasoner helped clarify their meaning and relations. Concepts such as `seamless' and `paradigm' may be interpreted in different ways depending on the person, but assigning their meaning in an ontology disambiguates them through definition-by-relations. Paradigm is defined as a Directive Information Content Entity (ICE) with continuant parts of Approach, Method, and Principle, which are all subclasses of Directive ICE, i.e., information content that prescribes some entity.

% (Listing~\ref{lst:Paradigm})
% \begin{lstlisting}[language=Manchester-syntax,
% label=lst:Paradigm,
% caption=\small{\textbf{Paradigm} is a subclass of \textbf{`Directive ICE'} equivalent to:}]
%     'Directive Information Content Entity'
% and ('has continuant part' some (Approach and Method and Principle))
% \end{lstlisting}

 Instead of defining `Seamless Digital Engineering' as a distinct concept, it needs to be decomposed into related concepts that fit appropriately as subclasses in the class hierarchy. Therefore, `Seamless Digital Engineering Paradigm' (Listing~\ref{lst:SDEP}) is defined as a subclass of Paradigm, although it may later be revised to be a subclass of Engineering Paradigm or similar when sibling classes are identified and defined.

\begin{lstlisting}[language=Manchester-syntax,
label=lst:SDEP,
caption=\small{\textbf{`Seamless Digital Engineering Paradigm'} is a subclass of \textbf{Paradigm} equivalent to:}]
    Paradigm
and ('has continuant part' some 'Digital Engineering Approach')
and ('is about' some Correct-by-Construction)
and (prescribes some 'Seamless Digital Engineering Environment')
\end{lstlisting}

Digital Engineering Approach is simply an Engineering Approach that prescribes some Act of Digital Engineering. Act of Digital Engineering is an Act of Engineering that has a Digital Artifact participant and an Authoritative Source of Truth as its object. Detailed definitions of these DE concepts are omitted here, and may be found in publications by the INCOSE DEIX Ontology Working Group. The essential concepts of the Seamless Digital Engineering Paradigm concept here are Correct-by-Construction (Listing~\ref{lst:CbC}) described by \citet{ManfredBroy_JanosSztipanovits2018} and the Seamless Digital Engineering Environment (Listing~\ref{lst:SDEE}).

\begin{lstlisting}[language=Manchester-syntax,
label=lst:CbC,
caption=\small{\textbf{Correct-by-Construction} is a subclass of \textbf{`Assurance Goal'} equivalent to:}]
    'Assurance Goal'
and ('is concretized by' some 'Integration Process')
and ('is concretized by' some 'Loss of Error')
and (prescribes some 'High-Integrity Level')
and (prescribes some 'Process Outcome')
\end{lstlisting}

Correct-by-Construction is modeled as an Assurance Goal, a descendant subclass of Directive ICE, suitable for referential use in system architecture models. The natural language definition (\texttt{skos:definition}) is derived from \citet{sztipanovits2012}: ``An Assurance Goal to eliminate Integration Process-time errors emerging from undesirable cross-layer interactions''. As a natural language elucidation, its component parts correspond to the ontology axiom above (Listing~\ref{lst:CbC}). High-Integrity Level is defined according to ISO/IEC/IEEE 15026 \citep{ISO15026-1} and is modeled as an Ordinal Measurement ICE from the CCO.

Seamless Digital Engineering Environment (Listing~\ref{lst:SDEE}) is an Engineered System, indicating it is designed, built, and validated to satisfy enumerated Stakeholder Needs. This acquisition approach contrasts with those commonly used for enterprise IT infrastructure supporting Digital Engineering Environments today, characterized by a series of ``build-or-buy'' decisions that typically lack the basis of an enterprise system architecture model. Rather than relying on third-party assessments that constitute a Trusted Computing Base, a Seamless Digital Engineering Environment relies on the independently-verifiable Seamless Quality Claim (Listing~\ref{lst:SeamlessQualityClaim}) of a Trustworthy Computing Base whose Trustworthiness is supported by a Complete Assurance Case Report defined by ISO/IEC/IEEE 15026 \citep{ISO15026-1} and provided as output of an Act of Assurance Evaluation defined by ISO/IEC 15408 \citep{ISO15408-1}.

\begin{lstlisting}[language=Manchester-syntax,
label=lst:SDEE,
caption=\small{\textbf{`Seamless Digital Engineering Environment'} is a subclass of \textbf{`Engineered System'} equivalent to:}]
    'Digital Engineering Environment'
and 'Engineered System'
and ('is carrier of' some ('High-Integrity Level Claim' and 'Seamless Quality Claim'))
and ('has member part' some 'Trustworthy Computing Base')
\end{lstlisting}

Seamless Quality Claim (Listing~\ref{lst:SeamlessQualityClaim}) is based on the concept of Claim defined by ISO/IEC/IEEE 15026 \citep{ISO15026-1}, which prescribes some Attribute, Condition, and Uncertainty, and is a descendant subclass of Constraint for use in a System Architecture Model. Seamless Quality Claim is the top-level Quality Claim for a Seamless Digital Engineering Environment, and links together the seamless composite Quality and Quality-in-Use concepts.

\begin{lstlisting}[language=Manchester-syntax,
label=lst:SeamlessQualityClaim,
caption=\small{\textbf{`Seamless Quality Claim'} is a subclass of \textbf{`Quality Claim'} equivalent to:}]
    'Quality Claim'
and (prescribes some 'Seamless Integration')
and (prescribes some 'Seamless Interaction Capability')
and (prescribes some 'Seamless Quality-in-Use')
\end{lstlisting}

Seamless Quality Claim is split among related composite qualities or Key Quality Attributes, and these results show how we disambiguate the term `seamless' using the SQuaRE quality \citep{ISO25010} and quality-in-use models \citep{ISO25019}. Seamless Integration (Listing~\ref{lst:SI}) is `described by' a Quality Characteristic and modeled as a Product Capability, a realizable entity, which relates Quality Sub-characteristics from SQuaRE to a correct Integration Process. Seamless Integration is realized by an Act of Formal Verification and specifically depends on Seamless Interface(s) (Listing~\ref{lst:SeamlessInterface}), defined as ``An Interface prescribed by a System Architecture Model which is the object of an Act of Formal Verification that produces a Proof Certificate of its Product Functional Correctness'' --- a SQuaRE Quality Sub-characteristic of Production Functional Suitability.

\begin{lstlisting}[language=Manchester-syntax,
label=lst:SI,
caption=\small{\textbf{`Seamless Integration'} is a subclass of \textbf{`Product Capability'} equivalent to:}]
    'Product Capability'
and ('has realization' some 'Act of Formal Verification')
and ('has continuant part' some 
        ('Product Analysability'
     and 'Product Faultlessness'
     and 'Product Functional Correctness'
     and 'Product Integrity'
     and 'Product Safe Integration'))
and ('specifically depends on' some 'Seamless Interface')
\end{lstlisting}

\begin{lstlisting}[language=Manchester-syntax,
label=lst:SeamlessInterface,
caption=\small{\textbf{`Seamless Interface'} is a subclass of \textbf{`Interface'}, a subclass of `Information Bearing Artifact', and is equivalent to:}]
    Interface
and ('has continuant part' some 'Proof Certificate')
and ('prescribed by' some 'System Architecture Model')
and ('is object of' some 'Act of Formal Verification')
\end{lstlisting}

Seamless Interaction Capability (Listing~\ref{lst:SIC}) relates to Product Interaction Capability with additional Product Functional Suitability sub-characteristics, and specifically depends on `Seamless Integration'. This concept captures the intuitive notion when a Human-Computer Interface (HCI) is seamless from the point-of-view of the Operator, and complements `Seamless Integration' which applies to Interfaces throughout the Seamless Digital Engineering Environment that the Operator may not interact with directly. Seamless Interaction Capability specifically depends on successful Seamless Integration.

\begin{lstlisting}[language=Manchester-syntax,
label=lst:SIC,
caption=\small{\textbf{`Seamless Interaction Capability'} is a subclass of \textbf{`Product Capability'} equivalent to:}]
    'Product Interaction Capability'
and ('has continuant part' some
        ('Product Compatibility'
     and 'Product Functional Appropriateness'
     and 'Product Functional Completeness'))
and ('specifically depends on' some 'Seamless Integration')
\end{lstlisting}

Seamless Quality-in-Use (Listing~\ref{lst:SQIU}) cannot be formally proven as it relies on the Operator's experience, but as with any Product Quality-in-Use, it should be traced to Stakeholder Needs in an appropriate Authoritative Source of Truth. When products claim to provide a `seamless experience', it is this Seamless Quality-in-Use comprising the ISO 25019 \citep{ISO25019} Qualities-in-Use, Experience and Trustworthiness (subclasses of Acceptability), and Suitability and Usability (subclasses of Beneficialness). Trustworthiness (Listing~\ref{lst:Trustworthiness}) is further decomposed to relate to the evaluation assurance concepts introduced above, and is therefore evidence-based.

\begin{lstlisting}[language=Manchester-syntax,
label=lst:SQIU,
caption=\small{\textbf{`Seamless Quality-in-Use'} is a subclass of \textbf{`Quality-in-Use'} equivalent to:}]
    Quality-in-Use
and ('has continuant part' some 
        (Experience
     and Suitability
     and Trustworthiness
     and Usability))
and ('specifically depends on' some
        ('Seamless Integration'
     and 'Seamless Interaction Capability'))
\end{lstlisting}

\begin{lstlisting}[language=Manchester-syntax,
label=lst:Trustworthiness,
caption=\small{\textbf{Trustworthiness} is a subclass of \textbf{`Acceptability'} equivalent to:}]
    Acceptability
and ('specifically depends on' some 
        ('Complete Assurance Case Report'
    and ('is carrier of' some 'Trustworthiness Quality Claim')))
\end{lstlisting}

Concepts from ISO/IEC/IEEE 15026 \citep{ISO15026-1,ISO15026-2,ISO15026-4} and ISO/IEC 15408 (``Common Criteria'') \citep{ISO15408-1,ISO15408-5} were carefully defined within the ontological framework. These concepts are essentially divided between Information Content Entities and their Information Bearing Artifact/Entity counterparts, supported by risk-related concepts defined as subclasses of Change, Stasis, Effect, and disposition. The ontology may be used to develop Complete Assurance Case Reports using Information Parts such as Evidence Items that conform to the standard. More complete systems engineering domain ontology development, including the many inputs/outputs/processes from ISO/IEC/IEEE 15288 \citep{ISO15288} will further enhance the usefulness of the ontology.

% \begin{figure}[b]
%     \centering
%     % \includegraphics[scale=0.8]{Figures/metamodel.pdf}
%     \caption{Some figure}
%     \label{fig:}
% \end{figure}

% \clearpage
% \section{Discussion} 
% \label{sec:discussion}

% \clearpage
\section{Conclusions and Future Work}
\label{sec:conclusions}

Ontology development in the systems engineering and digital engineering knowledge domains is ongoing internally among multiple groups, including INCOSE DEIX WG, AIAA Digital Engineering Integration Committee, INCOSE GfSE CASCaDE, OMG Model-Based Acquisition Working Group, and other entities, varying in ontological frameworks used and level of rigor. Final results from those working groups are forthcoming. We have presented concepts from the open-source Seamless Digital Engineering Ontology \citep{SDEontologyRepo} which includes over 500 classes and over 150 axioms based on 30 existing international standards and INCOSE technical products, in an attempt to align with the latest knowledge of the field. Definition of concepts in Digital Engineering is ongoing, and this ontology was developed in tandem with the smaller-scoped INCOSE DEIX Ontology to be published soon. This paper presented concepts that helped elucidate Seamless Digital Engineering, focusing on aspects of trustworthy systems, high assurance and high integrity, and product quality of the Digital Engineering Environment as an Engineered System. Future work includes a definition of object properties, especially those in the systems engineering domain, which was deferred to focus on the class hierarchy and use of existing object properties.

Harmonization of the concepts remains a difficult area, as terms in the literature may not be used consistently, or more importantly, do not match the base ontological definitions provided by BFO and CCO. Salient examples include different types of models which are often defined as ``representations'', but in CCO, a Representational ICE is distinct from an Artifact Model, which is a kind of Directive ICE. In these cases, the intent of the canonical sources was interpreted within the CCO framework, and the natural language definition was adjusted to fit the ontological axioms. Another challenge was disambiguation, including overloaded terms such as Authoritative Source of Truth, which required months of deliberations by the first author and INCOSE DEIX Ontology WG. We found that splitting concepts up according to the BFO and CCO framework was useful in clarifying their meaning and making further use of the concepts in relations. One such example is Traceability defined by ISO/IEC/IEEE 15288 \citep{ISO15288} which has been provisionally disambiguated into four related concepts: Traceability Measurement ICE, Traceability Relation, Traceability Observation Artifact Function, and Act of Establishing Traceability. The concept of System is defined as an object aggregate, but its subtler meanings and uses may require many revisions to incorporate it usefully in the CCO framework and to be useful to a variety of fields. Such challenging ontology development requires repeated discussions among subject matter experts, but the effort is worth it to produce standardized machine- and human-readable artifacts that have been machine-verified for logical consistency. We expect this harmonization work to continue as stakeholders in our field recognize the importance and power of ontologies in solving DE challenges.

Ontological definitions of Seamless DE concepts are offered based on existing and up-to-date international standards on product quality, but work remains to further define the quality characteristics using additional concepts from the domains pictured in Fig.~\ref{fig:SDEImportHierarchy}. The Trustworthiness Quality-in-Use Characteristic was selected for its importance in Seamless Quality-in-Use and its relation to existing standards in the systems engineering domain that define concepts in evaluation assurance. `Seamless' has been an appealing word to use when describing products, but it had not been given a precise meaning that would assist in MBSE-based development of clean-slate engineered systems. By following the SQuaRE model, we prepare the Seamless DE grand challenge for further development using Quality Need Expressions and Quality Requirement Expressions that shall be incorporated into a DE Environment reference architecture. Within this ontological framework, sub-disciplines of DE, such as digital requirements engineering and digital quality engineering, may be defined and distinguished from their traditional counterparts, potentially furthering our understanding and development of Digital Engineering.

%\clearpage % needs to be on a separate page
% \bibliography{references.bib}
\hypertarget{sec:References}{}
\bibliographystyle{plainnat}
\bibliography{main}

% \clearpage
% \appendix

% \section*{Appendix}
% \renewcommand{\thefigure}{A.\arabic{figure}} % Redefine figure numbering for the appendix
% \setcounter{figure}{0} % Reset figure counter for the appendix
% \label{sec: appendix}

%\clearpage % needs to be on a separate page

\section{Biography}

% NOTE: spacing is still not exact here compared to the Word document

\vspace{-3\baselineskip}

\noindent
\begin{center}
\begin{tabularx}{\textwidth}{p{1.25in} X}
\\[10pt]
\centering
\includegraphics[width=1in, valign=t]{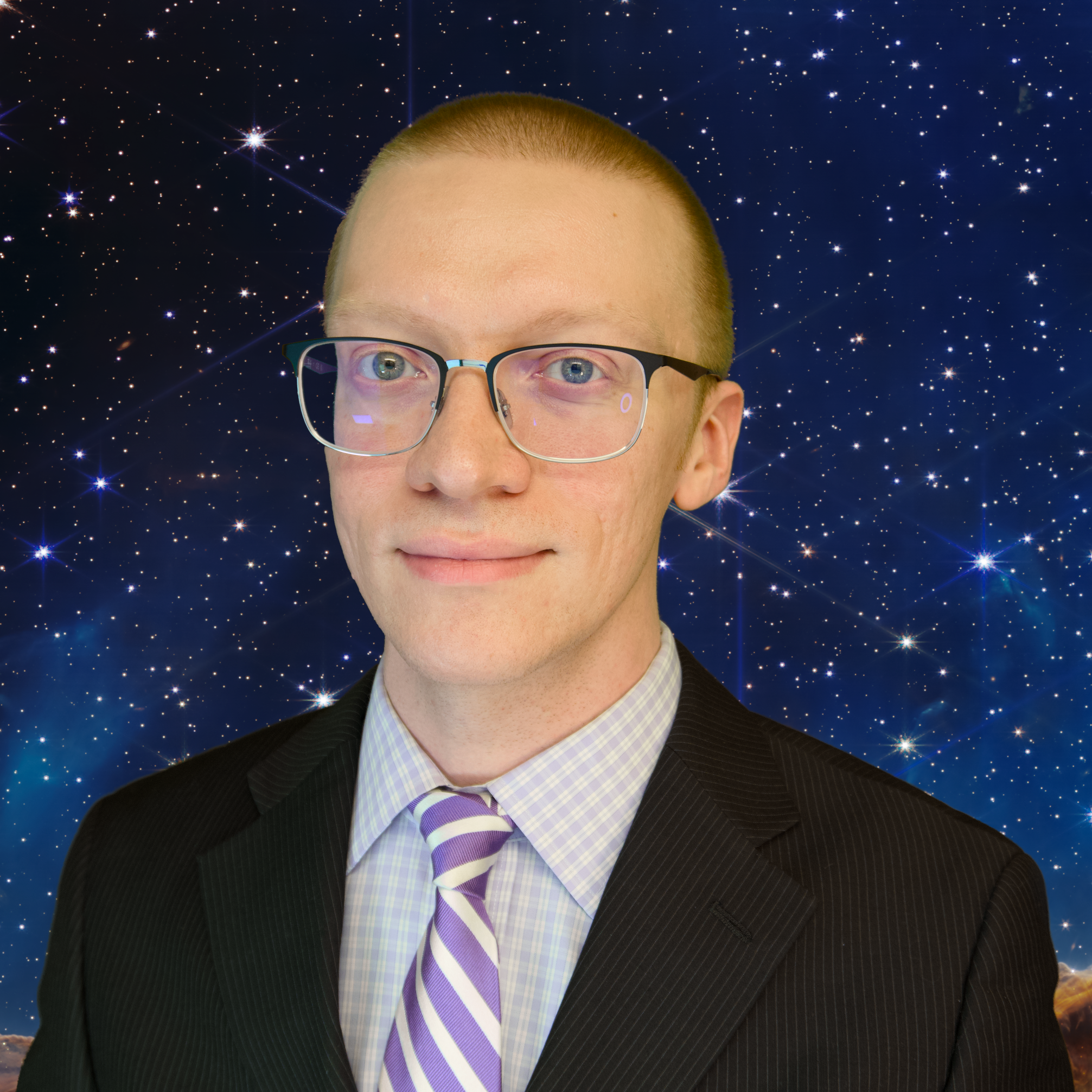}
& \textbf{James S. Wheaton} is a Ph.D. Candidate in Systems Engineering at Colorado State University in the Herber Research Group. His dissertation is titled “Bootstrapping a Trustworthy and Seamless Digital Engineering Appliance”. He is an INCOSE member and lead contributor in the INCOSE DEIX Ontology Working Group. James holds a B.S. in Mechanical Engineering from Purdue University (2011) and his professional background is in software engineering of ecommerce, big data, AI, and blockchain systems. James interned at NASA Jet Propulsion Laboratory from 2023-2024, where he developed system architectures and requirements for two Mars Sample Return projects.
\\
\\
\centering
\includegraphics[width=1in, valign=t]{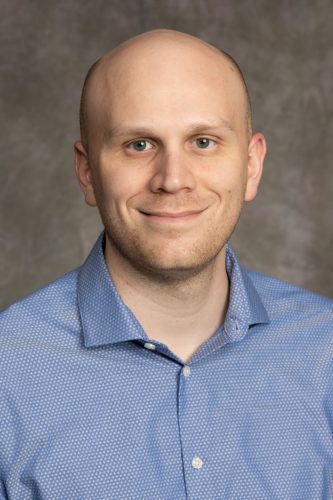}
& \textbf{Daniel R. Herber} is an Assistant Professor in the Department of Systems Engineering at Colorado State University in Fort Collins, CO, USA.
His research interests and projects have been in design optimization, model-based systems engineering, system architecture, digital engineering, dynamics and control, and combined physical and control system design (control co-design), frequently collaborating with academia, industry, and government laboratories.
His work has involved several application domains, including energy, aerospace, defense, and software systems. He teaches courses in model-based systems engineering, system architecture, controls, and optimization. He is a member of INCOSE, ASME, and AIAA.
He studied at the University of Illinois at Urbana-Champaign, earning his B.S. (2011) in General Engineering and his M.S. (2014) and Ph.D. (2017) in Systems and Entrepreneurial Engineering. He held a postdoctoral position (2018-2019) with the NSF ERC for Power Optimization for Electro-Thermal Systems (POETS).
\end{tabularx}
\end{center}

% DRH: old
% is an Assistant Professor in the Systems Engineering Department at Colorado State University.
% His research interests are in the areas of computational design, model-based systems engineering, digital engineering, design optimization, and combined physical and control system design (control co-design) concentrated around the development of novel theory and tools for integrated design methods conducive to emerging and dynamic engineering systems. His work has involved several engineering application domains taking a systems perspective on development and design, including the design of offshore wind/wave energy systems, thrust reversers, carbon capture systems combined with thermal storage and natural gas plants, thermal management networks for aircraft, and strain-actuated solar arrays for reorienting spacecraft.
% He studied at the University of Illinois at Urbana-Champaign, earning his B.S. in General Engineering in 2011 and his M.S. and Ph.D. in Systems and Entrepreneurial Engineering in 2014 and 2017, respectively. He held a postdoctoral position (2018-2019) with the NSF Engineering Research Center for Power Optimization for Electro-Thermal Systems (POETS).

\end{document}